\documentclass[prl, aps, twocolumn, groupedaddress,showpacs]{revtex4}
\usepackage{eqnarray}
\usepackage[english]{babel} 
\usepackage[english]{layout}
\usepackage{amsmath,amsfonts,amssymb}
\usepackage{graphicx}
\usepackage{float}
\usepackage{color}
\usepackage{braket}
\usepackage{version}
\usepackage{array,multirow,makecell}

\begin{document}

\title{Photonic Versus Electronic Quantum Anomalous Hall Effect}
\author{O. Bleu, D. D. Solnyshkov, G. Malpuech}
\affiliation{Institut Pascal, PHOTON-N2, Clermont Auvergne University, CNRS, 4 avenue Blaise Pascal, 63178 Aubi\`{e}re Cedex, France.} 

\begin{abstract}
We derive the diagram of the topological phases accessible within a generic Hamiltonian describing quantum anomalous Hall effect for photons and electrons in honeycomb lattices in presence of a Zeeman field and Spin-Orbit Coupling (SOC). The two cases differ crucially by the winding number of their SOC, which is 1 for the Rashba SOC of electrons, and 2 for the photon SOC induced by the energy splitting between the TE and TM modes. As a consequence, the two models exhibit opposite Chern numbers $\pm 2$ at low field. Moreover, the photonic system shows a topological transition absent in the electronic case. If the photonic states are mixed with excitonic resonances to form interacting exciton-polaritons, the effective Zeeman field  can be induced and controlled by a circularly polarized pump. This new feature allows an all-optical control of the topological phase transitions. 
\end{abstract}

\maketitle
The discovery of the quantum Hall effect \cite{PhysRevLett.45.494} and its explanation in terms of topology \cite{thouless1982quantized,PhysRevLett.51.2167} have refreshed the interest to the band theory in condensed matter physics leading to the definition of a new class of insulators \cite{RevModPhys.83.1057,RevModPhys.82.3045}. They include quantum anomalous Hall (QAH) phase \cite{com} with broken time reversal (TR) symmetry \cite{haldane1988model,PhysRevLett.101.146802,PhysRevB.82.161414} (also called Chern or  $\mathbb{Z}$ insulators) and Quantum Spin Hall (QSH or $\mathbb{Z}_2$) Topological Insulators with conserved TR  symmetry \cite{bernevig2006quantum,kane2005z,konig2007quantum}. 
The QSH effect was initially predicted to occur in honeycomb lattices because of the intrinsic Spin-Orbit Coupling (SOC) of the atoms forming the lattice, whereas the extrinsic Rashba SOC is detrimental for QSH \cite{kane2005z}. On the other hand, the classical anomalous Hall effect is now known to arise from a combination of extrinsic Rashba SOC and of an effective Zeeman field \cite{MacDonald2010}. In a 2D lattice with Dirac cones it leads to the formation of a QAH phase, for which the intrinsic SOC is detrimental \cite{PhysRevB.85.115439,zhang2011quantum,ren2016topological}. 
In the large Rashba SOC limit, this description was found to converge towards an extended Haldane model \cite{PhysRevB.85.115439}. Another field, which has considerably grown these last years, is the emulation of such topological insulators with different types of particles, such as fermions (either charged, as electrons in nanocrystals \cite{kalesaki2014dirac,beugeling2015topological}, or neutral, such as fermionic atoms in optical lattices \cite{jotzu2014experimental,PhysRevLett.116.225305}) and bosons (atoms, photons, or mixed light-matter quasiparticles) \cite{PeanoPRX2015,yuen2016plexciton,peano2015topological,rechtsman2013photonic,hafezi2011robust,hafezi2013imaging,
khanikaev2013photonic,cheng2016robust,CRAS2016}. The  main advantage of artificial analogs is the possibility to tune the parameters \cite{schmidt2015optomechanical}, to obtain inaccessible regimes, and to measure quantities out of reach in the original systems. These analogs also call for their own applications, beyond those of the originals. Photonic systems have indeed allowed the first demonstration the QAHE \cite{Soljacic2009,Soljacic2014}, later implemented in electronic \cite{chang2013experimental} and atomic systems \cite{aidelsburger2015measuring}. They have allowed the realization of topological bands with high Chern numbers ($C_n$) \cite{PhysRevLett.115.253901}, making possible to work with superpositions of chiral edge states. From an applied point of view, they open the way to non-reciprocal photonic transport, highly desirable to implement logical photonic circuits. On the other hand, the study of interacting particles in artificial topologically non-trivial bands could allow direct measurements of Laughlin wavefunctions (WFs) \cite{PhysRevLett.108.206809} and give access to a wide variety of strongly interacting fermionic \cite{wang2014classification} and bosonic phases \cite{peotta2015superfluidity}. In that framework, the use of interacting photons, such as cavity polaritons, for which high quality 2D lattices have been realized \cite{PhysRevLett.112.116402,Milicevic2015}, showing collective properties, such as macroscopic quantum coherence and superfluidity \cite{PhysRevLett.93.166401}, could allow to study the behaviour of bosonic spinor quantum fluids \cite{carusotto2013quantum,shelykh2009polariton} in topologically non-trivial bands. 
In photonics, a Rashba-type SOC cannot be implemented for symmetry reasons, but another effective in-plane SOC is induced by the energy splitting between the TE and TM modes. In planar cavities, the related effective magnetic field has  a winding number 2 (instead of 1 for Rashba). It is at the origin of a very large variety of spin-related effects, such as the optical spin Hall effect \cite{PhysRevLett.95.136601, leyder2007observation}, half-integer topological defects \cite{hivet2012half,dominici2014vortex}, Berry phase for photons \cite{Shelykh2009}, and the generation of topologically protected spin currents in polaritonic molecules \cite{PhysRevX.5.011034}. The combination of a TE-TM SOC and a Zeeman field in a honeycomb lattice has indeed been found to yield a QAH phase \cite{nalitov2015polariton,CRAS2016,PhysRevB.93.085438,KarzigPRX2015,LiewPRB2015,PhysRevB.93.104303,gulevich2016kagome}, and the related model represents a generalization of the seminal Haldane-Raghu proposal \cite{haldane2008possible} of photonic topological insulator, recovered for large TE-TM SOC. 

 In this manuscript, we demonstrate the role played by the winding number of the SOC on the QAH phases. We establish the complete phase diagram for both the photonic and electronic graphene. In addition to opposite $C_n$ in the low-field limit, we find the photonic case to be more complex, showing a topological phase transition absent in the electronic system. We then propose a realistic experimental scheme to observe this transition based on spin-anisotropic interactions in a macro-occupied cavity polariton mode. We consider a driven-dissipative model and demonstrate an all-optical control of these topological transitions and  of the propagation direction of the edge modes. One of the striking features is that the topological inversion can be achieved at non-zero values of the TR-symmetry breaking term, allowing chirality control by weak modulation of the pump intensity.

 \textit{Phase diagram of the photonic and electronic QAH.} We recall the linear tight-binding Hamiltonian of a honeycomb lattice in presence of Zeeman splitting and SOC of Rashba \cite{PhysRevB.79.165442} and photonic type respectively \cite{PhysRevLett.114.026803}. It is a $4$ by $4$ matrix written on the basis $(\Psi_A^+,\Psi_A^-,\Psi_B^+,\Psi_B^-)^T$, where $A$ and $B$ stand for the lattice atom type and $\pm$ for the particle spin:
\begin{eqnarray}
H_{k_{i}}=\begin{pmatrix}
\Delta \sigma_z&F_{k,i} \\
F_{k,i}^{\dagger} & \Delta \sigma_z
\end{pmatrix} 
,\quad F_{k,i}=-\begin{pmatrix}
f_{k}J&f_{k,i}^+\lambda_{i} \\
f_{k,i}^- \lambda_{i}&f_kJ 
\end{pmatrix}.
\end{eqnarray}
$J$ is the tunnelling coefficient between nearest neighbour micropillars (A/B). $\Delta$ is the Zeeman splitting. $\lambda_i$ ($i=e,p$) are  the magnitude for the Rashba (electronic) and TE-TM (photonic) induced SOC respectively \cite{suppl}.  The complex coefficients $f_k$ and $f_{k,i}^{\pm}$ are defined by:
\begin{eqnarray}
f_{k} =\sum_{j=1}^3e^{(-i\textbf{kd}_{\phi_j})},~ f_{k,e}^{\pm}&=\pm& \sum_{j=1}^3 e^{(-i[\textbf{kd}_{\phi_j}\mp \phi_j])} \\ f_{k,p}^{\pm}&=&\sum_{j=1}^3 e^{(-i[\textbf{kd}_{\phi_j}\mp 2\phi_j])} \nonumber
\end{eqnarray}
where $\textbf{d}_{\phi_j}$ are the links between nearest neighbour pillars (atoms) and $\phi_j=2\pi(j-1)/3$ their angle with respect to the horizontal axis. Qualitatively, the crucially different $\phi$ dependencies of the tunneling $f_{k,i}^{\pm}$ are due to the different winding numbers of the Rashba and TE-TM effective fields in the bare 2D systems.

 \begin{figure}[tbp]
 \begin{center}
 \includegraphics[scale=0.39]{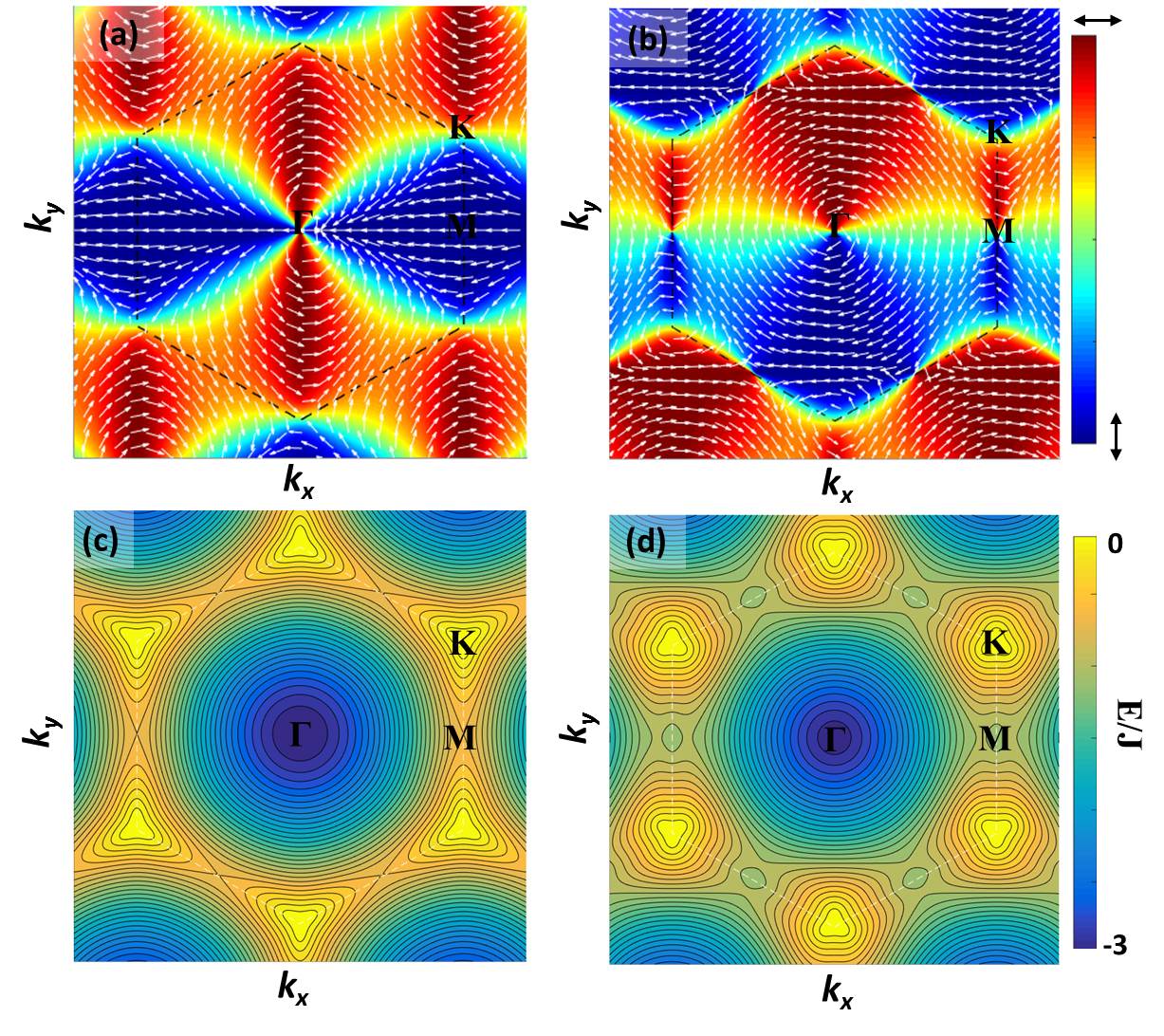}
 \caption{ (Color online) (a)-(b) Spin polarization textures in presence of TE-TM and Rashba SOC respectively (second branch). White arrows -- the in-plane spin projection. (c)-(d) Dispersions for TE-TM and Rashba SOC. The trigonal warping appears in different directions. ($\Delta=0$, $\lambda_{e,p}=0.2J$).}
  \label{pola}
  \end{center}
 \end{figure}
Without Zeeman field ($\Delta=0$), the diagonalization of these two Hamiltonians gives 4 branches of dispersion. Near $K$ and $K'$ points, two branches split, and two others intersect, giving rise to a so-called trigonal warping effect, namely the appearance of three extra crossing points (see  (Fig.~1(c,d) and Fig.~3(a)). 
The differences between the two Hamitonians are clearly visible on the panels of Fig. 1 which show a 2D view of the 2nd branch spin polarizations (a,b) and energies (c,d). On the panels (a,b),  we see the difference of the in-plane winding number around $\Gamma$ ($w_{\Gamma,e}=1$ for Rashba and $w_{\Gamma,p}=2$ for TE-TM SOC). Around K points, the TE-TM SOC texture becomes Dresselhaus-like with a winding $w_{K,p}=-1$ whereas Rashba remains Rashba with $w_{K,e}=1$. In each case, the winding numbers around the $K$ and $K'$ points have the same sign and add to give $\pm2$ $C_n$ for the electronic and photonic case respectively when TR is broken. On the panels (c,d), one can clearly observe the formation of small triangles near the Dirac points, the vertices of these triangles corresponding to the crossing points with the third energy bands. We can observe that the vertices are oriented along the $K-K'$ direction for TE-TM SOC and rotated by 60$^\circ$ ($K-\Gamma$ direction) for the Rashba SOC case, a small detail, which has crucial consequences for the topological phase diagram. 
The topological character of these Hamiltonians with the appearance of the QAH effect has already been discussed by deriving an effective Hamiltonian close to the $K$ point in different limits for both the electronic \cite{PhysRevB.82.161414, PhysRevB.85.115439} and photonic cases \cite{CRAS2016,nalitov2015polariton}. However, the presence of other topological phase transitions due to additional degeneracies appearing in other points of the first Brillouin zone was not checked.

   \begin{figure}[tbp]
 \begin{center}
 \includegraphics[scale=0.42]{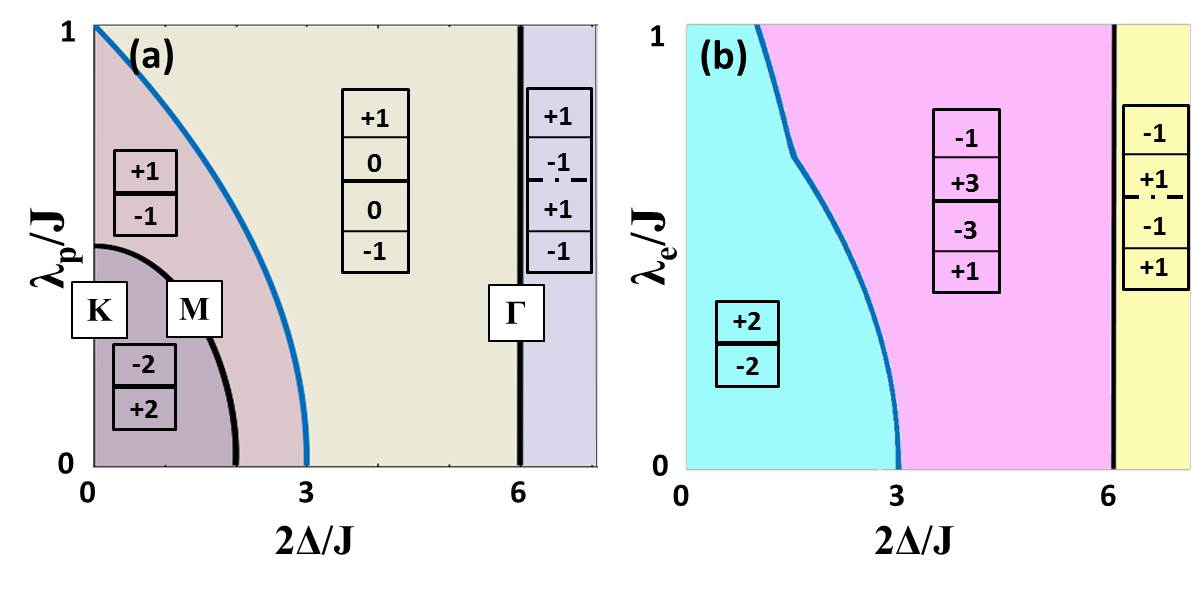}
 \caption{ (Color online) Phase diagrams (a) for the TE-TM SOC and (b) for the Rashba SOC with an applied field $\Delta$. Each phase is marked by the $C_n$ of the bands.}
  \label{phases}
  \end{center}
 \end{figure}
 
Figure 2 shows the diagram of topological phases of both models versus the SOC and Zeeman field strength.
 The different phases are characterized by the band $C_n$  that we calculate using the standard gauge-independent and stable technique of \cite{fukui2005chern}. We remind that change of $C_n$ is necessarily accompanied by gap closing. Obviously, these phase diagrams are symmetric with respect to $\Delta=0$ (with inverted signs of $C_n$ for the negative part). At low $\Delta$, both models are characterized by $C_n=\pm 2$. However, their $C_n$ signs are opposite due to the opposite winding of their SOC around K.
  \begin{figure}[bp]
 \begin{center}
 \includegraphics[scale=0.3]{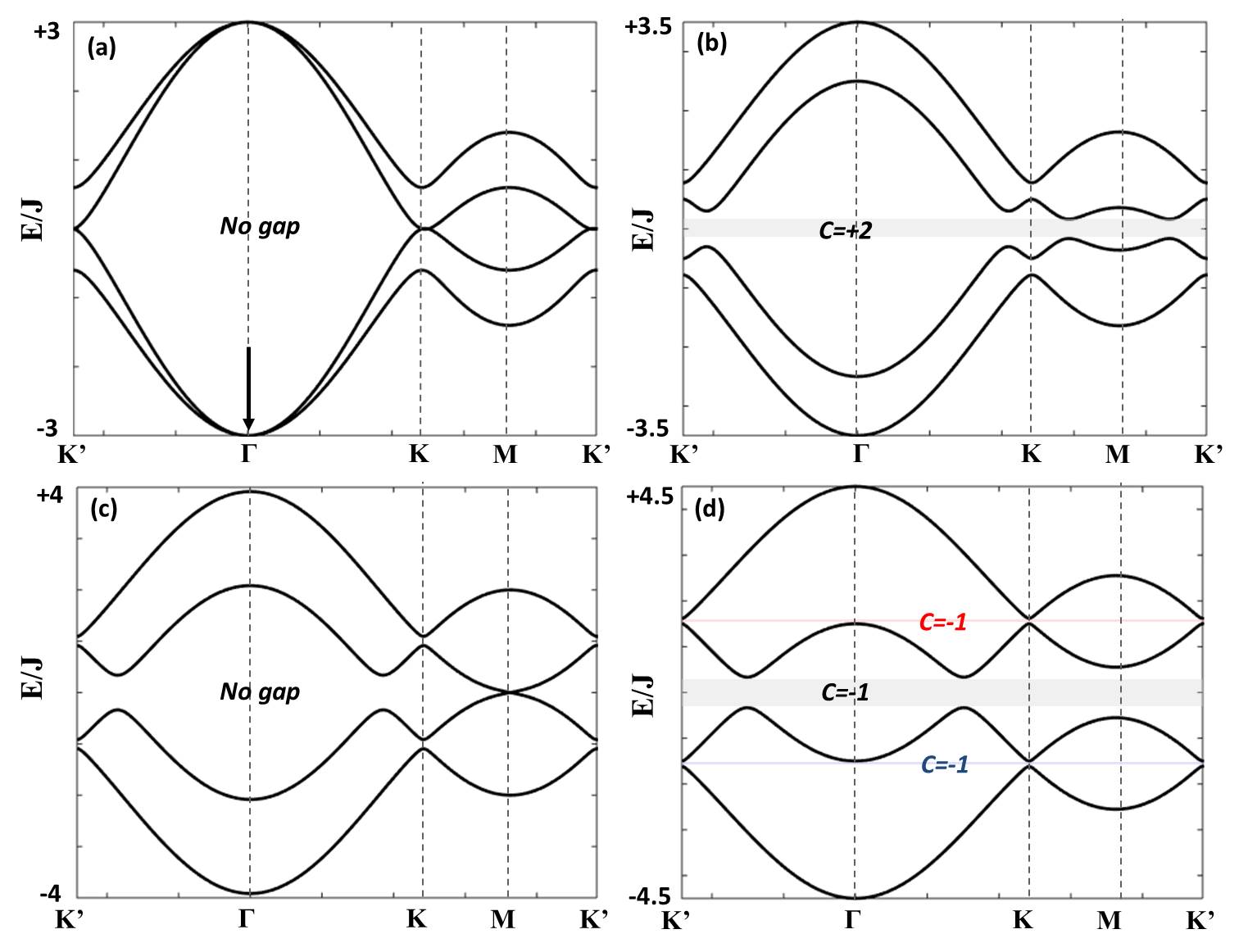}
 \caption{ (Color online) Dispersion of photonic graphene for different Zeeman field. (a) $\Delta=0$, (b) $\Delta=0.5J$, (c) $\Delta=\Delta_1$, (d) $\Delta=1.5J$. ($\lambda_p=0.2J$). The different gaps are shown in grey with the values of the associated $C_n$.}
  \end{center}
    \label{disp}
 \end{figure}

  Figure 3(b) shows the corresponding band structure for the photonic case, where the double peak structure around K and K', arising from the trigonal warping effect and responsible for the $C_n$ value, is clearly visible. Increasing either the SOC or the Zeeman field shifts these band extrema.  In the photonic case, the band extrema finally meet at the $M$ point, which makes the gap close, as shown on the figure 3(c). The critical Zeeman field value at which this transition takes place can be found analytically: $\Delta_1=\sqrt{J^2-4\lambda_p^2}$. Increasing the fields further leads to an immediate re-opening of the gap with the $C_n$ passing from +2 to -1 for the valence band. This case is shown on the figure 3(d), where the number of band extrema is twice smaller than on 3(b). This phase transition is entirely absent in the electronic case because of the different orientations of the trigonal warping.

Increasing the field even further leads to a second topological transition this time present in both models and associated with the opening of two additional gaps between the two lower and two upper branches (in the middle of the "conduction" band and of the "valence" band, correspondingly), as shown on the figure 3(d). This transition arises, when the minimum energy of the second branch at the $\Gamma$ point is equal to the maximal energy of the lowest band at the $K$ point, and thus the system of 2 bands (each containing 2 branches) is split into 4 bands (each containing a single branch). The corresponding transition in the photonic case occurs when the Zeeman splitting is: $\Delta_2=3(J^2-\lambda_p^2)/2J$. The last topological phase transition occurs when the middle gap closes at the $\Gamma$ point for $\Delta_3=3J$ and then reopens as a trivial gap, whereas the two other bandgaps are still topological.

\textit{All-optical control of topological phase transitions.} In what follows, we propose a practical way to implement the  photonic topological phases analyzed above.
We concentrate on the experimentally realistic configuration of a resonantly driven photonic (polaritonic) lattice \cite{PhysRevLett.112.116402,Milicevic2015}, including finite particle lifetime, without any applied magnetic field, and demonstrate the all-optical control of the band topology.  We show that the topologically trivial band structure becomes non-trivial under resonant circularly polarized pumping at the $\Gamma$ point of the dispersion. A self-induced topological gap opens in the dispersion of the elementary excitations. The tuning of the pump intensity allows to go through several topological transitions demonstrating the chirality inversion.

A coherent macro-occupied state of exciton-polaritons is usually created by resonant optical excitation. This regime is well described in the mean-field approximation \cite{PhysRevLett.93.166401,PhysRevB.77.045314}. We can derive the driven tight-binding  Gross-Pitaevskii equation in this honeycomb lattice for a homogeneous laser pump $F$ ($\hbar=1$).
\begin{eqnarray}
i\frac{\partial}{\partial t}\Psi_i= \sum_{j}H_{ij}(\textbf{k})\Psi_{j}+ F_{i}e^{i(\textbf{(k}_p.\textbf{r}-\omega_pt)}\nonumber
\\ +(\alpha_1\left|\Psi_i\right|^2+\alpha_2\left|\Psi_{i+(-1)^{(i+1)\mod 2}}\right|^2)\Psi_i
\label{GP-graphene1}
\end{eqnarray}
where ${i,j}=1..4$ correspond to the four WF components $(\Psi_A^+,\Psi_A^-,\Psi_B^+,\Psi_B^-)$. $H_{ij}$ are the matrix elements of the tight-binding Hamiltonian defined above (eq. 1) without the Zeeman term on the diagonal ($\Delta=0$). $\alpha_1$ and $\alpha_2$ are the interaction constants between particles with the same and opposite spins, respectively. For polaritons, the latter is suppressed \cite{Ciuti98} because it involves intermediate dark (biexciton) states, which are energetically far from the polariton states. Thus $ |\alpha_2| \ll \alpha_1 $ \cite{Renucci2005,Vladimirova2010} and we neglect it.
$F_{i}$ is the pump amplitude. In the following, we consider a homogeneous pump at $k=0$ (pumping beam perpendicular to the cavity plane), which implies that its amplitude on A and B pillars is the same. However, the spin projections $F_s^\sigma$ and $F_s^{-\sigma}$, determining the spin polarization of the pump, can be different ($s$ - sublattice, $\sigma$ - spin). The quasi-stationary driven solution has the same frequency and wavevector as the pump ($\Psi_{s}^{\sigma}=e^{i(\textbf{k}_p.\textbf{r}-\omega_pt)}\Psi_{p,s}^{\sigma}$) and satisfies the equations:
\begin{eqnarray}
&(\omega_p + i \gamma_p -\alpha_1 |\Psi_{p,s}^{\sigma}|^2 -\alpha_2 |\Psi_{p,s}^{-\sigma}|^2) \Psi_{p,s}^{\sigma} \nonumber \\
&+f_{k_p}J \Psi_{p,-s}^{\sigma} +f_{k_p}^{\sigma}\lambda_p \Psi_{p,-s}^{-\sigma}=F_s^{\sigma} \label{stat1}
\end{eqnarray}
where $\omega_p$ is the frequency of the pump mode.
 $\gamma_p$ is the linewidth related to polariton lifetime ($\tau_p$), which allows to take the dissipation into account. The tight-binding terms ($f_{k_p}$,$f_{k_p}^{\sigma}$) of the polariton graphene induce a coupling between the sublattices and polarizations. Eq. \eqref{stat1} is written for an arbitrary pump wave vector $k_p$. In the following, we consider a pump resonant with the energy of the bare lower polariton dispersion branch in the $\Gamma$ point ($\omega_p=-f_{\Gamma}J=-3J$ and $k_p=0$), marked with an arrow in Fig.~3(a) which implies the stability of the elementary excitations. 
We compute the dispersion of the elementary excitations using the standard WF of a weak perturbation ($|\textbf{u}|$,$|\textbf{v}|\ll|\vec{\Phi}_p |$):
\begin{equation}
\vec{\Phi}=e^{i(\textbf{k}_p.r-\omega_p t)}(\vec{\Phi}_p+\textbf{u}e^{i(\textbf{k}.r-\omega t)}+\textbf{v}^*e^{-i(\textbf{k}.r-\omega^* t)})
\end{equation}
where $\vec{\Phi}_p=(\Psi_{p,A}^+,\Psi_{p,A}^-,\Psi_{p,B}^+,\Psi_{p,B}^-)^T$ , $\textbf{u}$ and $\textbf{v}$ are vectors of the form $(u_A^+,u_A^-,u_B^+,u_B^-)^T$  \cite{suppl}.
 \begin{figure}[tb]
 \begin{center}
 \includegraphics[scale=0.4]{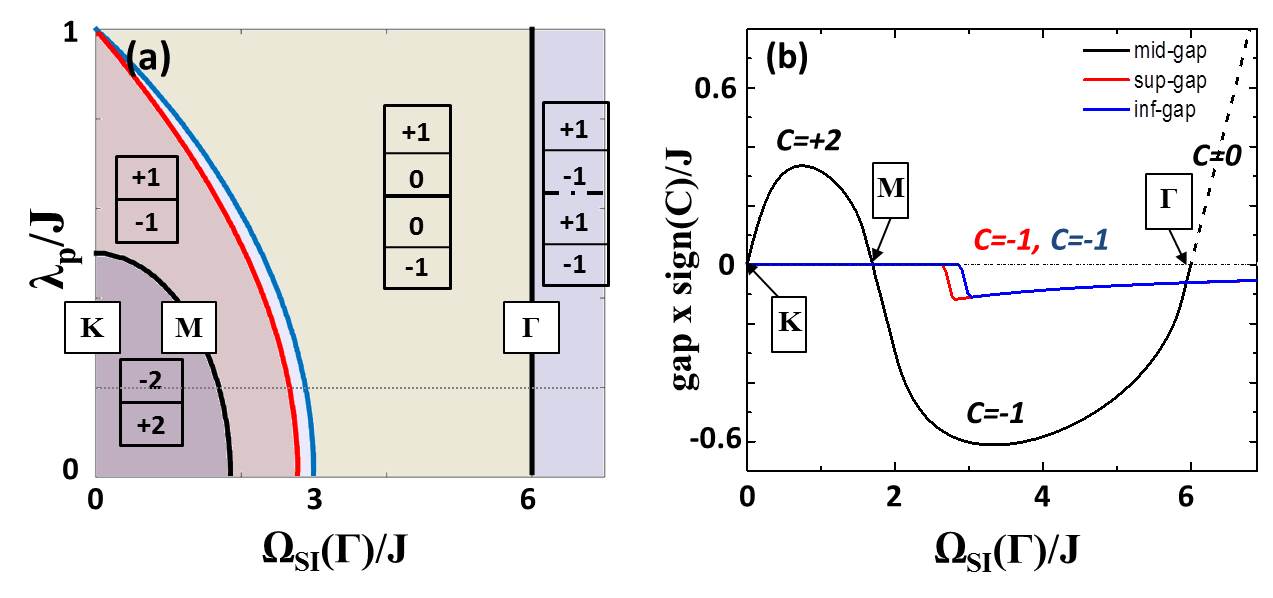}
 \caption{(Color online) (a) Topological phase diagram in the resonant pump regime versus the TE-TM SOC and SIZ. (b) Gap sizes and sign evolution along a path of constant SOC $\lambda_p=0.2J$ (dashed line on (a)). The red and blue curves correspond to the opening of additional gaps.}
  \label{nonlin}
  \end{center}
 \end{figure}
 
A circular pump induces circularly polarized macro-occupied state ($n^-=0$), and $n=n^+=n_A^+ +n_B^+=|\Psi_{p,A}^{+}|^2+|\Psi_{p,B}^{+}|^2$. Combined with spin anisotropic interactions, it leads to a Self-Induced Zeeman (SIZ) splitting which breaks TR symmetry. A simple analytical formula of the $k$-dependent SIZ splitting between the two lower branches is obtained for $\lambda_p =0$: 
\begin{equation}
\Omega_{SI}=\omega_p+|f_k|+\sqrt{(\omega_p+|f_k|J-2 \alpha_1 n_{A/B}^+)^2- (\alpha_1 n_{A/B}^+)^2}\nonumber
\end{equation}
One of the key differences with respect to the magnetic field induced Zeeman field is the SIZ dependence on the wavevectors and energies of the bare modes. This dependence has already been shown to lead to the inversion of the effective field sign (and thus the inversion of the topology) when both applied and SIZ fields are present in a Bose-Einstein condensate \cite{PhysRevB.93.085438}.

The figure 4(a) shows the diagram of topological phases under resonant pumping (versus the SIZ) which is  quite similar to the one under magnetic field. A method to compute the $C_n$ of the Bogoliubov modes has been developed in \cite{Shindou2013}. The procedure we use is detailed in the supplementary \cite{suppl}.
The only difference with respect to the linear case concerns the opening of the two additional gaps which does not take place at the same pumping values, because of the difference between the SIZ fields in the upper and lower bands. The figure 4(b) shows  the magnitude of the different gaps multiplied by the sign of the $C_n$ of the valence band ($C=\sum_{i=1}^n C_n$) \cite{PhysRevLett.71.3697} for a given value of the SOC, a quantity highly relevant experimentally. In \cite{PhysRevLett.112.116402,Milicevic2015} $J$ is of the order or 0.3 meV, whereas the mode linewidth is of the order of 0.05 meV. Band gaps  of the order of 0.2 $J$ should be observable. The SIZ magnitude shown on the x-axis (below 1.5 meV) is compatible with the experimentally accessible values. So in practice the topological transition is observable together with the specific dispersion of the edge states in the different phases which are presented in \cite{suppl}.
We note that the emergence of topological effects driven by interactions in bosonic systems has already been reported, such as Berry curvature in a Lieb lattice for atomic condensates \cite{DiLiberto2016} and topological Bogoliubov edge modes in two different driven schemes based on Kagome lattices \cite{PhysRevB.93.020502,peano2015topological} with scalar particles. 

 \begin{figure}[tb]
 \begin{center}
 \includegraphics[scale=0.32]{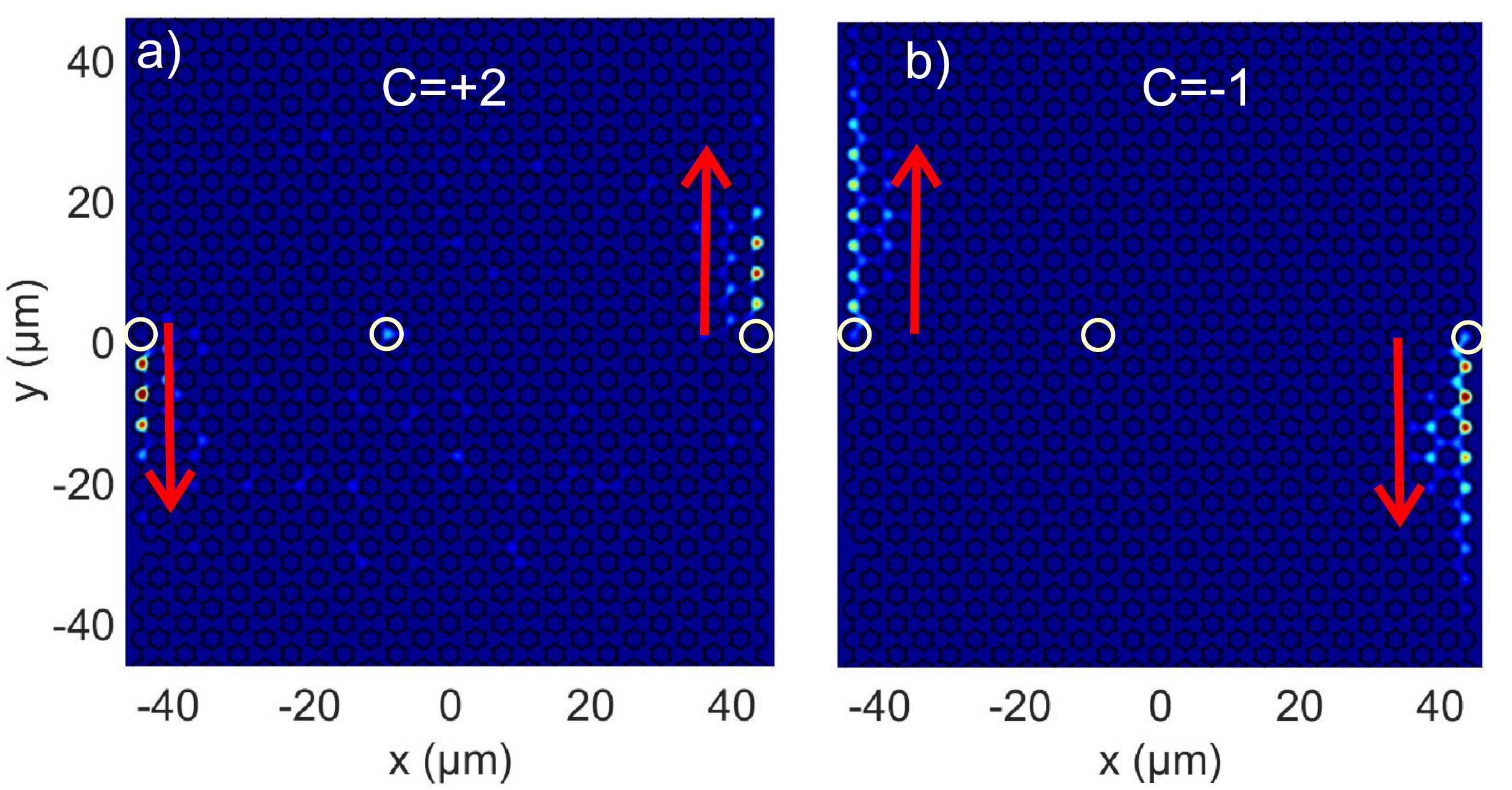}
 \caption{ (Color online) Calculated images of emission from the surface states (a) $\Omega_{SI}=0.3$ meV, $C=2$  (b) $\Omega_{SI}=0.6$ meV, C=-1. Arrows mark the propagation direction. }
  \label{surfacestates}
  \end{center}
 \end{figure}

To confirm our analytical predictions and support the observability in a realistic pump-probe experiment (see sketch in \cite{suppl}), we perform a full numerical simulation beyond the tight-binding or Bogoliubov approximations. We solve the spinor Gross-Pitaevskii equation for polaritons with quasi-resonant pumping:
\begin{eqnarray}
& i\hbar \frac{{\partial \psi _ \pm  }}
{{\partial t}}  =  - \frac{{\hbar ^2 }}
{{2m}}\Delta \psi _ \pm +\alpha_1\left|\psi_\pm\right|^2\psi_\pm    - \frac{{i\hbar }}
{{2\tau }}\psi _ \pm  + P_{0}^+e^{-i\omega t}\\
& + U\psi _ \pm + \beta {\left( {\frac{\partial }{{\partial x}} \mp i\frac{\partial }{{\partial y}}} \right)^2}{\psi _ \mp }  +\sum_{j}P_{j}^- e^{ { - \frac{{\left( {t - t_0 } \right)^2 }}
{{\tau _0^2 }}- \frac{{\left( {{\mathbf{r}} - {\mathbf{r}}_j } \right)^2 }}
{{\sigma ^2 }}-{i\omega t} } } \nonumber
\end{eqnarray}
where ${\psi_+(\mathbf{r},t), \psi_-(\mathbf{r},t)}$ are the two circular components of the WF, $m=5\times10^{-5}m_{el}$ is the polariton mass, $\tau=30$ ps the lifetime, $U$ is the lattice potential. The main pumping term $P_{0+}$ is circular polarized ($\sigma^+$) and spatially homogeneous, while the 3 pulsed probes are $\sigma^-$ and localized on 3 pillars (circles). The results (filtered by energy and polarization) are shown in Fig. \ref{surfacestates}. As compared with the previously analyzed \cite{nalitov2015polariton,PhysRevB.93.085438} $C=2$ case (a), a larger gap of the $C=-1$ phase (b) demonstrates a better edge protection, a longer propagation distance, and an inverted direction, all achieved by modulating the pump intensity.

\textit{Conclusions.}  We bridge the gap between two classes of physical systems where the QAH effect takes place, showing the crucial role of the SOC winding. In the photonic case, we show that the phases achieved, their  topological nature and topological transitions can be controlled by optically induced collective phenomena. Our results show that photonic implementations of topological systems are not only of practical interest, but also bring new physics directly observable in real-space optical emission.

\bibliography{biblio} 

\begin{thebibliography}{72}
\bibitem[{com()}]{com}
\bibinfo{note}{One should note that the QAH can also be called equivalently
  $\mathbb{Z}$ topological insulator or Chern insulator. In the present
  manuscript we have have chosen the label QAH which is mostly used for the
  corresponding electronic systems. Actually, a similar twofold terminology
  exists for the Quantum spin Hall phase which can be called $\mathbb{Z}_2$
  topological insulator.}
\end{thebibliography}

\section{Supplemental material}
In this supplemental material, we first reintroduce the  TE-TM SOC.  Then, we provide details concerning the second part of the main text on the all-optical control of topological phase transitions. 
\subsection{Optical spin-orbit coupling}
In the main text, we introduce two kind of SOC $\lambda_e$ and $\lambda_p$ for electrons and polaritons respectively. We choose this notation to make clearer the comparison between the two cases. Indeed, in our precedent works on polariton honeycomb lattices, we used the notation $\lambda_p=\delta J$ \cite{PhysRevLett.114.026803,nalitov2015polariton,PhysRevB.93.085438}.
Taking into account the TE-TM splitting the tunneling coefficients are defined in the circular-polarization basis as:
\begin{eqnarray}
 \bra{A,\pm}H\ket{B,\pm}&=&-J  \\ \bra{A,\pm}H\ket{B,\mp}&=&-\lambda_p e^{-2i\phi_j} \nonumber
 \end{eqnarray}
 $J$ is the tunneling  coefficient without spin inversion, like in conventional graphene. The SOC coefficient $\lambda_p$ is defined by: $\lambda_p=\delta J=(J_L-J_T)/2$, where $J_L$ and $J_T$ are the tunneling coefficients for the longitudinally and transversally-polarized polaritons respectively.
The  difference of phase between Rashba ($e^{\pm i\phi}$) and TE-TM ($e^{\pm i2\phi}$) terms comes from the different winding number between the two in plane SOC.

 \subsection{Weak excitations dispersions in the resonant pump regime}

In this section we present the derivation of the Bogoliubov excitation of a resonantly pumped interacting photon system. The weak perturbation of the pumped macro-occupied state reads:

 \begin{equation}
\vec{\Phi}=e^{i(\textbf{k}_p.r-\omega_p t)}(\vec{\Phi}_p+\textbf{u}e^{i(\textbf{k}.r-\omega t)}+\textbf{v}^*e^{-i(\textbf{k}.r-\omega^* t)})
\end{equation}
where $\vec{\Phi}_p=(\Psi_{p,A}^+,\Psi_{p,A}^-,\Psi_{p,B}^+,\Psi_{p,B}^-)^T$ , $\textbf{u}$ and $\textbf{v}$ are vectors of the form $(u_A^+,u_A^-,u_B^+,u_B^-)^T$ too.  
Indeed because of the non-linear term of the GP equation, the Bloch state characterized by a wave-vector $k$ and frequency $\omega$ is coupled to its complex conjugated, namely the wave with a  wave-vector $-k$ and frequency $-\omega$.
Then, inserting this wave function in the driven dissipative Gross-Pitaevskii equation (main text) and linearizing for  $\textbf{u}$ and $\textbf{v}$, we obtain the following matrix:

\begin{widetext}
\begin{scriptsize}
\begin{equation}
\setlength\arraycolsep{0.4pt}
M=\begin{pmatrix}
(d_A^+-\omega_p-i\gamma_p ) &\alpha_2\Psi_{p,A}^{-*}\Psi_{p,A}^{+}&-f_{k_p+k}J & -f_{k_p+k}^+\lambda_p& \alpha_1 \Psi_{p,A}^{+2}& \alpha_2 \Psi_{p,A}^{-}\Psi_{p,A}^+&0&0 \\

\alpha_2\Psi_{p,A}^{+*}\Psi_{p,A}^{-} &(d_A^--\omega_p-i\gamma_p)&-f_{k_p+k}^-\lambda_p & -f_{k_p+k}J &\alpha_2 \Psi_{p,A}^{-}\Psi_{p,A}^+ & \alpha_1 \Psi_{p,A}^{-2} &0&0\\

 -f_{k_p+k}J &-f_{k_p+k}^{-*}\lambda_p&(d_B^+-\omega_p-i\gamma_p)&\alpha_2\Psi_{p,B}^{-*}\Psi_{p,B}^{+}&0&0& \alpha_1 \Psi_{p,B}^{+2}& \alpha_2 \Psi_{p,B}^{-}\Psi_{p,B}^+\\
 
  -f_{k_p+k}^{+*}J &-f_{k_p+k}\lambda_p& \alpha_2\Psi_{p,B}^{-}\Psi_{p,B}^{+*} &(d_B^--\omega_p-i\gamma_p)&0&0& \alpha_2 \Psi_{p,B}^{-}\Psi_{p,B}^+&\alpha_1 \Psi_{p,B}^{-2} \\

-\alpha_1 \Psi_{p,A}^{+2*} & -\alpha_2 \Psi_{p,A}^{-*}\Psi_{p,A}^{+*}&0 &0 & (\omega_p-i\gamma_p-d_A^+)& -\alpha_2\Psi_{p,A}^{-}\Psi_{p,A}^{+*}&f_{k_p-k}^*J&f_{k_p-k}^{+*}\lambda_p\\

-\alpha_2 \Psi_{p,A}^{-*}\Psi_{p,A}^{+*}&-\alpha_1 \Psi_{p,A}^{-2*} &0 &0& -\alpha_2\Psi_{p,A}^{-*}\Psi_{p,A}^{+} & (\omega_p-i\gamma_p-d_A^-)&f_{k_p-k}^{-*}\lambda_p&f_{k_p-k}^{*} J\\

0& 0&-\alpha_1 \Psi_{p,B}^{+2*} &-\alpha_2 \Psi_{p,B}^{-*}\Psi_{p,B}^{+*}&f_{k_p-k}^{*} J & f_{k_p-k}^{-*} \lambda_p &(\omega_p-i\gamma_p-d_B^+)&-\alpha_2\Psi_{p,A}^{-}\Psi_{p,A}^{+*} \\

0& 0&-\alpha_2 \Psi_{p,B}^{-*}\Psi_{p,B}^{+*}&-\alpha_1 \Psi_{p,B}^{-2*} &f_{k_p-k}^{+*}\lambda_p & f_{k_p-k}^{*}  J &-\alpha_2\Psi_{p,B}^{-*}\Psi_{p,B}^{+} & (\omega_p-i\gamma_p-d_B^-)\\
\end{pmatrix}
\end{equation}

\end{scriptsize}
\end{widetext}
The diagonal elements are defined by:
\begin{equation}
d_s^\sigma=2\alpha_1|\Psi_{p,s}^\sigma|^2+\alpha_2|\Psi_{p,s}^{-\sigma}|^2
\end{equation}
The Bogoliubov eigenenergies and eigenvectors $(u_A^+,u_A^-,u_B^+,u_B^-,v_A^+,v_A^-,v_B^+,v_B^-)^T$ are finally obtained by diagonalizing this 8 by 8 matrix. 

In the expression for the self-induced field $\Omega_{SI}(\Gamma)/2=\sqrt{3}\alpha_1n/2$  the factor $1/2$ comes from the presence of two sublattices and the $\sqrt{3}$ appears from resonant pumping, as compared with a blue shift of an equilibrium condensate $\mu=\alpha n$.

The normalisation condition, requires for the Bogoliubov transformation to be canonical, namely to keep bogolons as bosons reads \cite{lifshitz1980statistical,pitaevskii2003bose}: 
\begin{eqnarray}
 \sum_{1\leq i\leq 4}|u_i|^2-|v_i|^2=1
\end{eqnarray}
where $i$ index labels the different $u_i$($v_i$) components of an eigenstate.

This condition physically signifies that the creation of one bogolon corresponds to the creation of a quanta of energy $\omega$.

\subsection{Chern numbers of Bogoliubov excitations}
The standard formula for the computation of the Chern number can be applied, but taking into account that bogolons are constituted by two Bloch waves of opposite wave vectors:

\begin{widetext}
\begin{eqnarray}
 C&=&\frac{1}{2i\pi} \iint\limits_{BZ} \nabla_\textbf{k} \times \bra{\Phi(\textbf{k})} \nabla_\textbf{k} \ket{\Phi(\textbf{k})}\mathrm{d}\textbf{k}\\
&=&\frac{1}{2i\pi} \iint\limits_{BZ}\nabla_\textbf{k} \times \bra{\textbf{u}(\textbf{k})} \nabla_\textbf{k} \ket{\textbf{u}(\textbf{k})}\mathrm{d}\textbf{k}+ \frac{1}{2i\pi} \iint\limits_{BZ}\nabla_\textbf{k} \times \bra{\textbf{v}(-\textbf{k})} \nabla_\textbf{k} \ket{\textbf{v}(-\textbf{k})}\mathrm{d}\textbf{k} \nonumber \\
&=&\frac{1}{2i\pi} \iint\limits_{BZ}\nabla_\textbf{k} \times \bra{\textbf{u}(\textbf{k})} \nabla_\textbf{k} \ket{\textbf{u}(\textbf{k})}\mathrm{d}\textbf{k}+ \frac{1}{2i\pi} \iint\limits_{-BZ}\nabla_{-\textbf{k}} \times \bra{\textbf{v}(\textbf{k})} \nabla_{-\textbf{k}} \ket{\textbf{v}(\textbf{k})}\mathrm{d}\textbf{k} \nonumber \\
&=& \frac{1}{2i\pi} \iint\limits_{BZ}\nabla_\textbf{k} \times \bra{\textbf{u}(\textbf{k})} \nabla_\textbf{k} \ket{\textbf{u}(\textbf{k})}\mathrm{d}\textbf{k}- \frac{1}{2i\pi} \iint\limits_{BZ}\nabla_\textbf{k} \times \bra{\textbf{v}(\textbf{k})} \nabla_\textbf{k} \ket{\textbf{v}(\textbf{k})}\mathrm{d}\textbf{k}
\end{eqnarray}
\end{widetext}
where $d\textbf{k}=\mathrm{d}k_x\mathrm{d}k_y$ and we drop the band index n for simplicity.
We can see that the integration of the $v$ part makes appear a minus sign because the integration takes place over an inverted Brillouin zone (BZ). This fact has been noticed in ref \cite{Shindou2013}, and is commonly used \cite{engelhardt2015topological,furukawa2015excitation,peano2015topological,DiLiberto2016}. It is typically formulated by introducing a matrix $\tau_z=\sigma_z \otimes 1\!\!1^4$ directly in the definition of the Berry connexion $\mathbf{A}=\bra{\Phi(\textbf{k})} \nabla_\textbf{k} \tau_z \ket{\Phi(\textbf{k})}$. 

 \subsection{Bogoliubov edge states}

 To demonstrate one-way edge states in tight-binding
approach, we derive a 8Nx8N Bogoliubov matrix for a polariton graphene stripe, consisting of N coupled infinite zig-zag chains following the procedure of Ref.\cite{PhysRevB.93.085438}. 
For this, we set a basis of Bogoliubov Bloch waves $(
u_{A/B,n}^{\pm}, v_{A/B,n}^{\pm})$ where n index numerates stripes, and $k_y$ is
the quasi-wavevector in the zigzag direction. The diagonal blocks describe coupling within one chain and are
derived in the same fashion as the M matrix in the previous section (2), coupling
between stripes is accounted for in subdiagonal blocks.

Figures 1(a,b) show the results of the band structure calculation for two different values of $\alpha_1n$ . The degree of localization on edges is calculated from the wave function densities on the edge chains $|\Psi_R|^2$ and $|\Psi_L|^2$ (left/right, see inset), and is shown with
colour, so that the edge states are blue and red. 

   \begin{figure}[H]
 \begin{center}
 \includegraphics[scale=0.6]{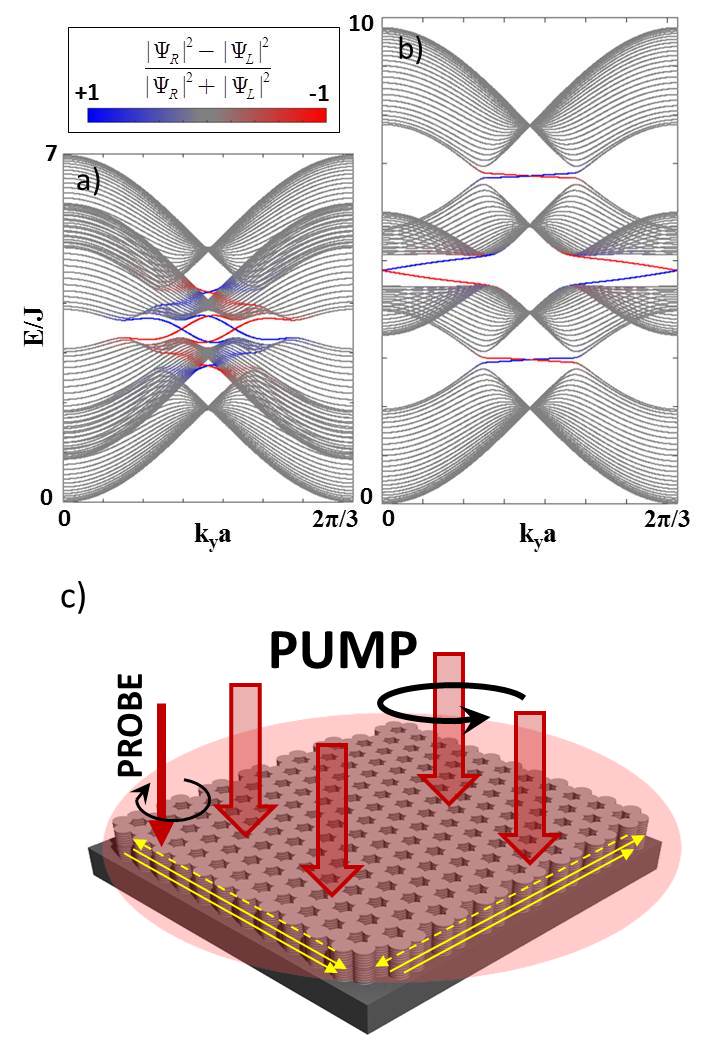}
 \caption{ (Color online) (a,b) Band structures of a graphene ribbon in two different phases. Blue and red colors refer to the states localized on the right and left edges. Parameters:  $\lambda_p=0.2J$ and (a) $\alpha_1n=1J$, (b) $\alpha_1n=4J$. (c) Real space sketch of the experimental setup. The yellow arrows represent the edge states when C=+2 (dashed ones when C=-1).}
  \end{center}
 \end{figure}
In Fig. 1(a), there is only one topological gap characterized by a Chern number $+2$ and hence there are two edge modes on each side of the ribbon. In Fig. 1(b), we can observe three topological gaps with the Chern number of the top and bottom bands being $\pm 1$ respectively.
 Each of them is characterized by the presence of only one edge mode on a given edge of the ribbon, and the group velocities of the modes are opposite to the previous phase: the chirality is controlled by the intensity of the pump. This inversion, associated with the change of the topological phase ($\left| C \right| = 2 \to 1$), is fundamentally different from the one of Ref. \cite{PhysRevB.93.085438}, observed for the same phase ($\left|C\right|=2$).

This optically-controlled transition allows to observe the inversion of chirality for weak modulations of a TR-symmetry breaking pump around a non-zero constant value, which can also possibly be used for amplification.
 The inversion of chirality of center gap edge states (Fig.~1(a,b)) should be observable in a pump-probe experiment as shown by the numerical simulation in the main text. A sketch of the experiment using a $\sigma^+$ and a $\sigma^-$ polarized lasers (the homogeneous pump and the localized probe) is presented on Fig.~1(c).
One should note that we can also obtain the inverted phases more conventionally by inverting the direction of the self-induced Zeeman field which is controlled by the circularity of the homogeneous pump.

\end{document}